\documentclass[conference]{IEEEtran}
\IEEEoverridecommandlockouts
\usepackage{cite}
\usepackage{amsmath,amssymb,amsfonts}
\usepackage{algorithmic}
\usepackage{graphicx}
\usepackage{textcomp}
\usepackage{xcolor}
\usepackage{hyperref}
\usepackage{multirow}
\def\BibTeX{{\rm B\kern-.05em{\sc i\kern-.025em b}\kern-.08em
    T\kern-.1667em\lower.7ex\hbox{E}\kern-.125emX}}
    
\usepackage{caption}
\usepackage{subcaption}

\usepackage{xcolor}

\begin{document}

\title{MIMA - Multifunctional IoT integrated \\Menstrual Aid
}

\author{
\IEEEauthorblockN{Amish Bibhu, Shreya Shivangi, and Sulagna Saha}
\IEEEauthorblockA{
    National Institute of Fashion Technology\\
    Bhubaneswar, Odisha 751024, India\\
    \{amish.bibhu, shreya.shivangi, sulagna.saha\}@nift.ac.in\\
    \{0000-0001-7708-110X, 0000-0003-3710-9571\}}
\and
\IEEEauthorblockN{Jyothish Kumar J, and Subhankar Mishra}
\IEEEauthorblockA{
    National Institute of Science Education and Research\\
    Bhubaneswar, Odisha 752050, India\\
    \{jyothishk.j, smishra\}@niser.ac.in\\
    \{0000-0001-7731-3281, 0000-0002-9910-7291\}}
}

\IEEEoverridecommandlockouts
\IEEEpubid{\makebox[\columnwidth]{978-1-6654-8356-8/22/\$31.00~\copyright2022 IEEE, DOI: 10.1109/COINS54846.2022.9854943 \hfill}
\hspace{\columnsep}\makebox[\columnwidth]{ }}

\maketitle

\IEEEpubidadjcol

\begin{abstract}
Menstruation is the monthly shedding of the endometrium lining of a woman's uterus. The average age when girls start menstruating is around the age of 12 years (menarche), and the cycle continues until they attain menopause (about the age of 51). Medical research and analysis in this field reveal that most women have to go through a painful cycle of abdominal cramps along with sanitary pad rashes, while painkillers or endurance ability are their go-to solution. Heat pads or hot water bags also help in pain reduction. Currently, the concept of period pants revolves around pad-free and hassle-free periods for women, whereas most women still prefer sanitary pads during their period cycle. 
MIMA aims at the development of IoT integrated smart, functional intimate wear for women that would help women comfort during menstruation by catering to issues of menstrual cramps, rashes, leakage and stains, malodor, etc. The proposed methodology has been implemented by referring to the online survey conducted from Indian women (17- 58 years old). MIMA can provide comfort during the menstruation cycle with IoT integrated Heat-Pad and functional alterations in the garment for a rash-free, anti-odor, and leak-proof period.
\end{abstract}


\section{Introduction}

Most women experience a regular menstrual cycle, but the fact that a large section of women faces multiple irregularities and issues during menstruation cannot be contradicted. Many adolescent girls have to deal with multiple cramps at an early age and are in constant fear of stains during their cycle. These are the symptoms of a menstrual irregularity known as Dysmenorrhea. Dysmenorrhea is a medical terminology referring to painful menstruation or menstrual cramps. It can be categorized into Primary and Secondary dysmenorrhea. Primary dysmenorrhea is the most commonly experienced recurrent menstrual cramps by women. The cramps and pain mostly start two to three days before the bleeding and can last up to the first three to four days of the cycle. The pain might range from mild to severe ache, mostly in the lower abdominal region. Recurring cramps and fatigue can also be felt in the lower back, hip, and thigh regions. Secondary dysmenorrhea could be an outcome of multiple problems with the reproductive organs, including Endometriosis, Pelvic inflammatory disease(PID), Adenomyosis, Fibroids, etc \cite{oladosu2018abdominal}. Constant use of sanitary pads is also the causal agent of rashes caused due to the continuous friction between the sensitive inner thigh region and the excess portion of the micro plastic present in pad wings. These all issues affect the work-life of women, their mood, productivity and also result in absenteeism from their workplace. Some of the major reasons for being absent from school are discussed in Table \ref{tab:reasons}

\begin{table}[ht]
\centering
\caption{Reason for being absent from school during menstruation among female students in Mehalmeda secondary school, Amhara region, Ethiopia, June 2013.\cite{gultie2014age}}
\label{tab:reasons}. 
\begin{tabular}{|l|l|l|}
\hline
\textbf{\#} & \textbf{Reason for being absent}        & \textbf{\% of students} \\ \hline
1             & Socio-culture belief                    & 0.4                              \\ \hline
2             & Shame                                   & 3.7                              \\ \hline
3             & shortage of water                       & 5.7                              \\ \hline
4             & Lack of disposal system                 & 8.5                              \\ \hline
5             & Fear of Leakage                         & 9.5                              \\ \hline
6             & Pain or discomfort                      & 10.5                             \\ \hline
7             & Lack of privacy for washing or cleaning & 13                               \\ \hline
\end{tabular}
\end{table}



MIMA is a high-waisted hipster intimate wear fabricated entirely with antibacterial finished cotton-elastane blend fabric. The alterations are done to the garment structure to achieve the critical functionalities like the improvisation of a leak-proof gusset along with the mini-pockets for accommodating the excess portions of the sanitary pad wings in order to prevent the rashes. The IoT intervention is done to incorporate a very slim and lightweight heat pad which would be placed in the inside pocket located in the abdominal region. The idea is to enable a comfortable range of heat to the abdominal region during cramps, and the control of the heat range lies in the hands of the user through a user-friendly app connected with Bluetooth. 

\begin{table}[h]
\caption{Issues faced during menstruation and Solution incorporated in the garment}
\label{tab:solutions}. 
\centering
\begin{tabular}{|l|p{0.4\linewidth}|p{0.4\linewidth}|}
\hline
\textbf{\#} & \textbf{Issues faced in Menstruation}                   & \textbf{Solution proposed in MIMA}                               \\ \hline
1 & Bacterial Infections & Anti-Bacterial Fabric   \\ \hline
2 & Menstrual Odor       & Anti-Bacterial Fabric   \\ \hline
3 & Menstrual Cramps     & IoT Powered Heating Pad \\ \hline
4           & Rashes due to Sanitary Pads                                         & Pockets integrated for Sanitary Pad Wings in Gusset area               \\ \hline
5           & Unavailability for the right size for girls under age of 12 & Integrating the same features in kids-wear with appropriate size chart \\ \hline
\end{tabular}
\end{table}

\section{Problem Statement}

\subsection{Menstruation}

Menstruation is a very critical and complex mechanism in a woman’s body that depends not just on reproductive organs but multiple hormones like oestrogen \& progesterone, food habits, emotional and physical well-being, heredity, and a lot more. It is the removal of the endometrium lining of the uterus through the vagina. The menstrual fluid consists of blood, cells from the uterus endometrium lining, and mucus. The average length of menstruation is usually between three to seven days but it can be of a one-day length to as long as eight days normally in a female. Pain-free menstruation and a regular menstrual cycle would require a perfect synchronization of all these factors.  \cite{thiyagarajan2020physiology}. Table \ref{tab:solutions} shows the solutions we are working on in MIMA to resolve the issues faced during menstruation. 

\subsection{Early onset of Menarche}
Girls around the age of 12 years usually hit puberty and attain Menarche, which refers to their first menstrual cycle. The Menarche marks the development of sexual characteristics. Ideally, the appropriate age to attain Menarche is expected to be from 12- 14 years of age yet now we can see significant deviations and multiple factors responsible for catalyzing the very early puberty, especially in girls of age 8-10 years. 
Several factors like intake of genetically engineered fruits and vegetables, presence of pesticides in food, increased poultry diet, etc., induce early puberty in kids. The sedentary lifestyle of children can lead to obesity and sometimes brings in a very stressful environment for themselves, which in turn triggers early puberty for girls.\cite{dharmarha2018study}


\subsection{Menstrual discomfort}


Early onset of menarche makes it difficult for adolescents to deal with severe cramps, sudden changes in their bodies, and terrible mood swings for more than a week. It is tough to deal with it and manage the school. Even for an adult, multiple cramps (both mild and severe pain), body aches, and headaches, along with fear of stains \& leakage, make it very tough for them to deal with. Researchers claim that around 84.1\% of women experience severe cramps during every menstruation resulting in absenteeism from the workplace and school\cite{grandi2012prevalence}.


During menstruation, one out of many problems the women face is menstruation malodor which makes them really conscious and worried at times. It is also a cause for discomfort due to which women prefer to be indoors during this time. The odor is an abnormal fishy smell caused due to the presence of bacteria \emph{Gardnerella vaginalis} in vaginal mucous along with the menstrual blood. The phenomenon is known as Bacterial vaginosis.\cite{mogilnicka2020microbiota}

\subsection{Menstrual Cramps and Remedies}
The most obvious query until now is why cramps? As we know, Menstruation is the monthly shedding of the endometrial lining of the uterus, and this shows some extra visceromotor reflexes resulting in cramps \cite{oladosu2018abdominal}. Many home remedies have been tried by women to get rid of the constant pain and cramps apart from pain killers, of which heat therapy and acupressure to the abdominal and hip region are their go-to remedies. Hot water bags, heating pads are proven conventional pain relief methods followed till today and are recommended by medical professionals. Research and surveys have been evident that superficial heat ranging from 40-45°C applied at a depth of 1-2 cm in the pelvic and abdomen region helps improve blood circulation. This eliminates the local blood fluid retention and diminishes congestion and swelling, reducing the pain caused by nerve congestion \cite{higgins2002quantifying}. The only drawback we see here is that the hot water bag could not be used in every place considering the availability of hot water. 

\section{Related Works}

\begin{table}[h]
\caption{Issues and existing solutions.\cite{peberdy2019study} \cite{healthshots}}\label{tab2}
\centering
\begin{tabular}{|p{0.2\linewidth}|p{0.7\linewidth}|}
\hline
\textbf{Issues}                                    & \textbf{Existing Solutions}                                     \\ \hline
\multirow{4}{*}{Cramp}                             & 1. Electric heating pads                                        \\ \cline{2-2} 
                                                   & 2. Hot Water Bag                                                \\ \cline{2-2} 
 & 3. Nua Heating Patches (An Eco-Friendly solution for cramp relief to be applied in the abdominal region) \\ \cline{2-2} 
                                                   & 4. Livia (Acupressure solution for cramp relief using impulses) \\ \hline
\multirow{4}{*}{Stain \& Leakage} & 1. Sanitary Pads                                                \\ \cline{2-2} 
                                                   & 2. Tampons                                                      \\ \cline{2-2} 
                                                   & 3. Menstrual Cup                                                \\ \cline{2-2} 
                                                   & 4. Period Pants                                                 \\ \hline
\multirow{3}{*}{Odor}                              & 1. Scented Sanitary Pads                                        \\ \cline{2-2} 
                                                   & 2. Anti-bacterial finish Intimate Wear                           \\ \cline{2-2} 
                                                   & 3. Essential Oil                                                \\ \hline
Rashes                                             & Rash cure with ointments                                        \\ \hline
Reminder                                           & Flow App and other Period tracker apps                          \\ \hline
\end{tabular}
\end{table}

 Period Pants have been launched by different brands to provide women a pad-free period whereas, in the case of Indian women, the majority of women of menstrual age use sanitary pads only\cite{smith2020national} \cite{choi2021use}. The working aims at achieving the high absorption capacity of multiple pads/ tampons secured with a leak-proof outer layer. From our survey, we found out that very few people are using period pants, and most of the research is not oriented toward the Indian market. The current research is being done for women/girls starting from the age of 12, considering the starting age for Menstrual cycles, but in the current scenario, girls have started to get their first menstrual cycle before the age of 12 also. \cite{ruble1982experience} There are multiple products and works to solve these problems individually, as shown in Table \ref{tab2} but there is a grey area identified in the development of a product with all-round features.
 
The period pants available currently, have three layers: 
\begin{enumerate}
    \item super-dry layer at the top 
    \item high absorbent layer in the middle capable of absorbing multiple numbers of sanitary pads/tampons.
    \item the last layer facing the main garment is the leak-proof layer.
\end{enumerate}

The main reason for the acceptance of period pants by women is the reusability and economic factor. Reusability makes it sustainable and eco-friendly, which is a key driving factor. The fact that a woman spends around \$10,000- \$15000 in her entire lifetime on purchasing menstrual aids and this switch to reusable period pants would save a lot, they are open to this option gladly, though the use in practicality is comparatively very low than the other sanitary aids\cite{phan2020acceptance}. There is quite a lot of motivation and awareness about period pants, but preferability is a lot less. A qualitative study outlines the values the surveyees place on the importance of proper sanitary infrastructure and menstrual hygiene education while in a refugee camp, it was a challenge to manage menstruation. Access to menstrual aids and their disposal is a major struggle. So, the dual use of reusable period pants and normal underwear was found to be very popularly preferable \cite{vanleeuwen2018exploring}. 

While we discuss different options of menstrual aids and talk about menstrual hygiene on multiple forums, there are still many rural households where women are not fortunate enough to be able to afford a pack of sanitary pads during menstruation, there are people who still use cloth pieces and have no option but to compromise their menstrual hygiene. \cite{kaur2018menstrual}


\section{Methodology}
\subsection{Data Selection and Sampling Technique}
The main focus of MIMA is to cater to the different problems faced by adolescent girls and women during menstruation, and to understand it briefly. We had discussions with our friends and family members about their discomfort. To get a more extensive and thorough understanding, we conducted a survey with questionnaires related to the irregularities they face during their menstrual cycle and understand their preferences and unapproached requirements in their intimate wear through simple questions formulated from our observations and reference of multiple papers. 

The structured online survey reached a sample size of 170  Indian women from different age groups and different professions who have attained menarche. The sampling technique was a purposive sampling technique where our data collection was in qualitative as well as quantitative format. 

\subsection{Data Collection and Analysis} 
The recorded data from the online questionnaire was analyzed using different statistical tools in Microsoft excel. The statistical and inferential analysis(as shown in Table \ref{tab:surveyresponse}) from the tools, tables, pivot tables, pivot charts and graphs used to test the reliability and validity of the questionnaire. The analysis of the responses directs us to our garment structure and functionalities.
\begin{table}[]
\caption{Data collected through our survey}\label{tab:surveyresponse}
\centering
\begin{tabular}{|l|p{0.2\linewidth}|p{0.62\linewidth}|}
\hline
\textbf{\#} &
  \textbf{Question Criteria} &
  \textbf{Response} \\ \hline
1 &
  Demography &
  The responses received, were from respondents of multiple age group \& professions and analysis shows that mostly our respondents were single students of age groups 17-21 years (48.24\%) \\ \hline
2 &
  Buying preference &
  Most of our respondents consider fabric (74.7\%) as the primary buying factor. Apart from this, comfort is of at most importance(85.3\%). We came to know that dark shade(64.1\%) of the intimate wear was also quite preferred during menstruation. \\ \hline
3 &
  Menarche &
  Most of our respondents have attained menarche within the age of 12-14 years (70\%) followed by girls who attained menarche before the age of 12(15.9\%). Also, 58.8\% of our sample size know someone in their friend and family who attained before the age of 12. \\ \hline
4 &
  Menstrual Aids &
  Our Analysis claims that 95.3\% of the respondents use sanitary pads in India and even if 56.5\% of them are aware of period pants, hardly anyone uses them. \\ \hline
5 &
  Menstrual Irregularities &
  Though 80\% of our respondents experience regular menstrual cycle while there are 57.1\% of the respondents who face issues during their regular or irregular cycle also. On categorization, we found that around 44.1\% of the respondents go through painful abdominal cramps post a few cycles and 34.1\% of them have to deal with multiple cramps. \\ \hline
6 &
  Cramp Remedy &
  Heat therapy is preferred for cramp relief by most of the women(58.8\%) where in most of them choose the medium range of heat (69.4\%). \\ \hline
7 &
  Preferred Intimate Wear &
  Most women prefer hipster intimate wear during their menstruation period and usual days followed by bikini style, \\ \hline
8 &
  Open towards New product &
  Through our survey, we came to know that around 76.5\% of our respondents are open to any new product catering to menstrual problems. \\ \hline
\end{tabular}
\end{table}

\subsection{Product Design}

Referring to the survey responses, we get an idea that fabric and comfort are decision-making factors when choosing intimate wear for most women. As the intimate area is very sensitive so, special preference is given to intimate wear made of natural fibers and a soft feel. Apart from these factors, the stretchability also could not be compromised. After going through multiple compositions of fabrics, considering the responses from our survey and their color preferences, we selected a navy blue fabric of composition 93\% cotton and 7\% elastane and 161 GSM(Gram per square meter) fabric with an anti-bacterial finish.

Data collection regarding the style preference of intimate wear on regular days as well as during the menstruation cycle gave us a clear idea that hipster is the most preferred style in both cases. So, our base pattern selected for pattern development is a basic hipster pattern on which we have made required alterations referring to the other requirements of our sample size.
The waist of the hipster is improvised to high-waist intimate wear, where an inside pocket is developed for accommodating a slim heat-pad at the abdominal region. Usually, the gusset area is a cut-2 panel, i.e., two identical gusset panels are cut, tacked to each other, and stitched from front to back. Considering our leak-proof feature, we have sandwiched a leak-proof layer in between the two gusset panels for the same, which prevents any leakage or stain-marks from the panties to the external bottom wear (example. denim pants, skirt, leggings, etc.). For incorporating the rash-free feature, we found that mainly the extra part of wings that come in contact with the inner thigh area of the women causes rashes, so we have developed tiny sanitary pad wing-shaped pockets tacked to the gusset area to accommodate the extra portion of the pad wings in order to minimize the chances of rashes in the respective area.

    

We developed the product in multiple samples stages and feedback was taken from women, and we made the required alterations to reach MIMA. 

\begin{figure}
     \centering
     \begin{subfigure}[H]{0.24\textwidth}
         \centering
         \includegraphics[width=\textwidth]{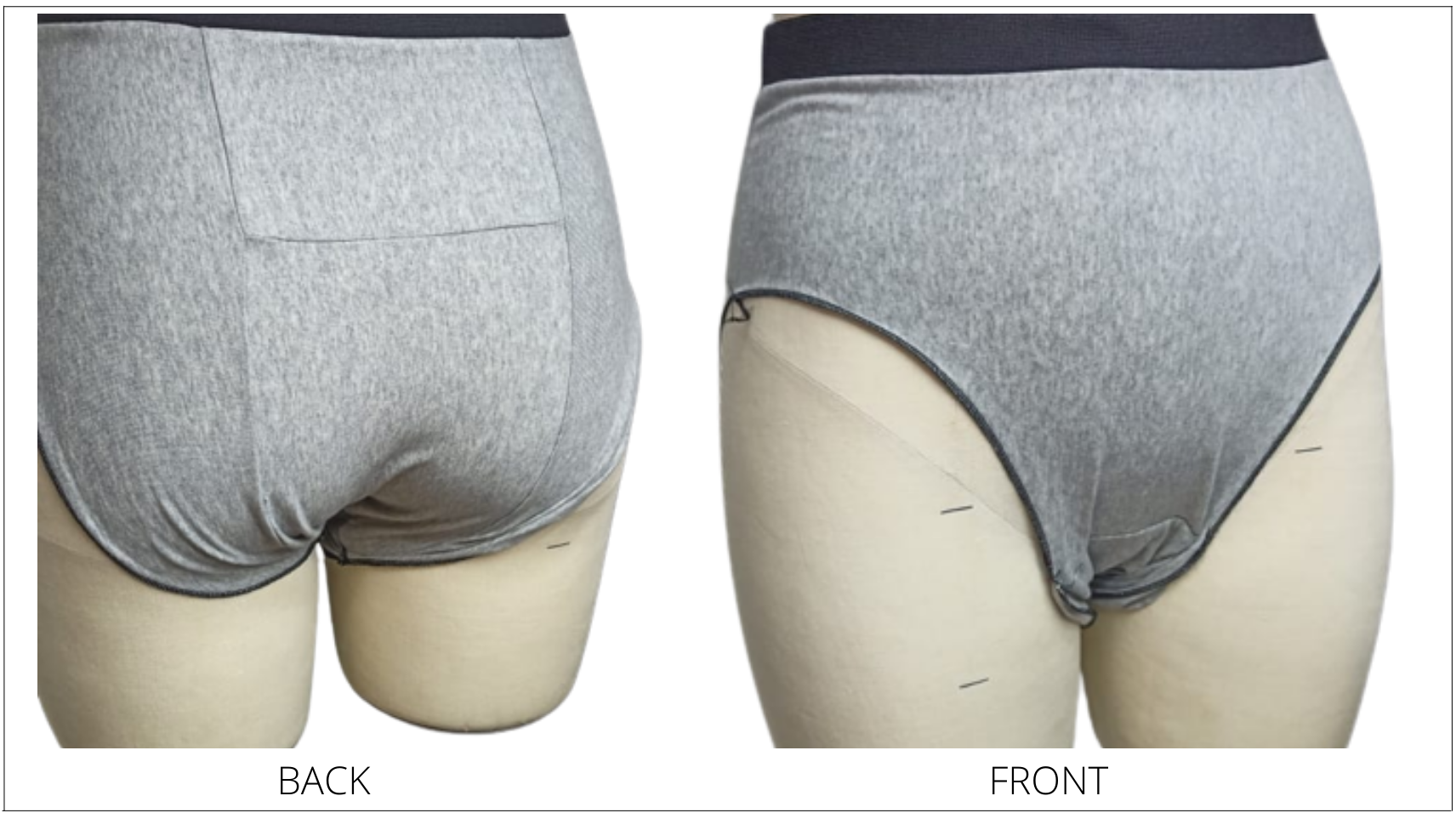}
         \caption{Proto Sample}
         \label{fig: Proto Sample}
     \end{subfigure}
     \hfill
     \begin{subfigure}[H]{0.24\textwidth}
         \centering
         \includegraphics[width=\textwidth]{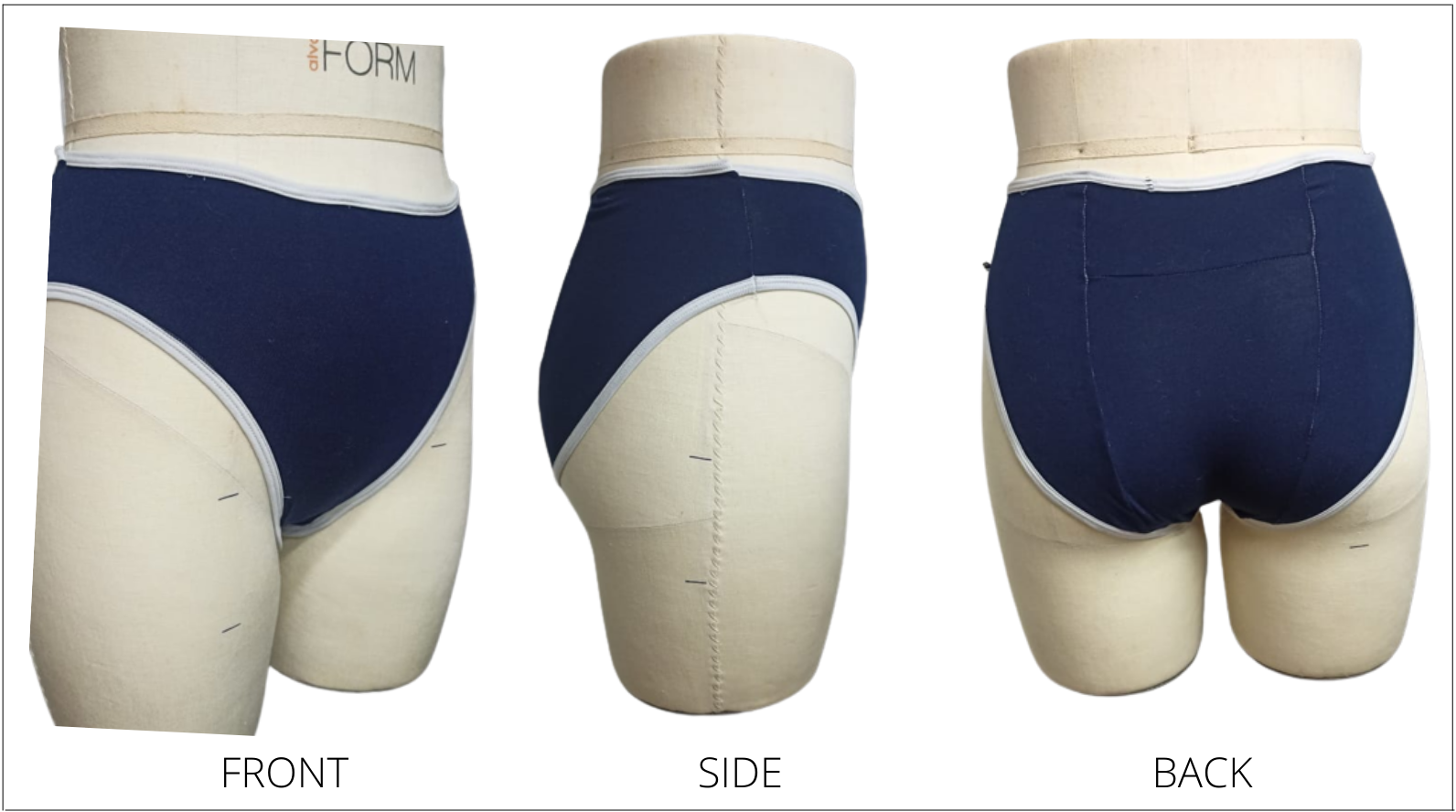}
         \caption{Fit Sample}
         \label{fig:Fit Sample}
     \end{subfigure}
     \hfill
     \begin{subfigure}[H]{0.24\textwidth}
         \centering
         \includegraphics[width=\textwidth]{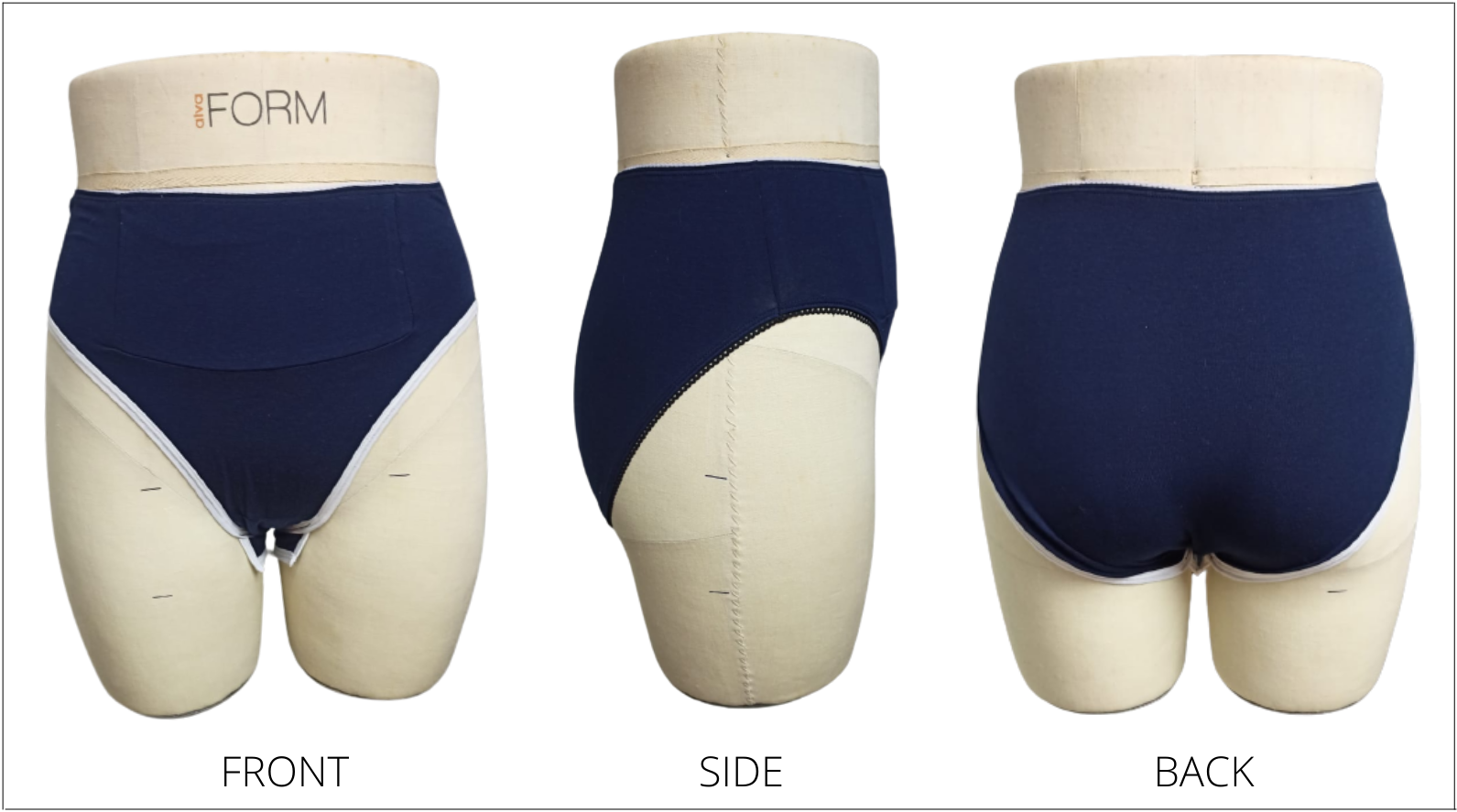}
         \caption{Final Prototype}
         \label{fig:Final design}
     \end{subfigure}
     \hfill
    \begin{subfigure}[H]{0.24\textwidth}
         \centering
         \includegraphics[width=\textwidth]{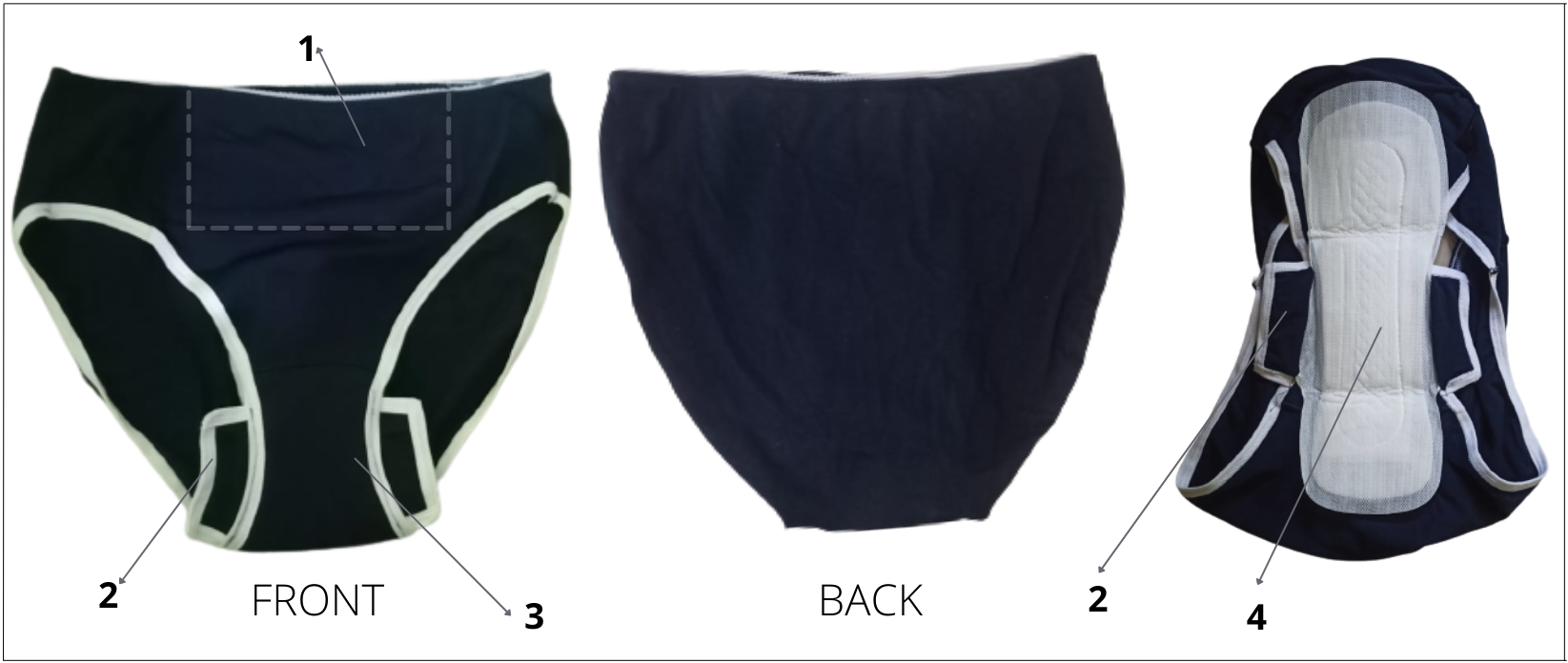}
         \caption{Features included in the garment}
         \label{fig:garmentexplained}
     \end{subfigure}
     \hfill
        \caption{Three stages of the garment sampling process}
        \label{fig:garment}
\end{figure}

\begin{enumerate}
    \item Our proto sample was made for a large size fit with normal knitted cotton fabric, and with its trial, we have identified some fit issues in the waist region. We realized that the waistband elastic and the style lines in the back we used in it could be modified to increase the comfort and complexity of the garment structure. (as shown in figure \ref{fig: Proto Sample}
     \item The fit sample was made with the main fabric (93\% cotton - 7\% elastane blend and with an anti-bacterial finish) and altered the waistband elastic to a thin \& light elastic band for the waist along with binding fabric stitched in the leg and waist region as shown in figure \ref{fig:Fit Sample}. After a few trials and feedback, we reached MIMA.
      \item MIMA is a navy-blue high-waisted hipster which fits perfectly into a Large size body type with the pocket easily accommodating the heat pad as well the wing pocket comfortably fitting in the inner thigh region, as shown in figure \ref{fig:Final design}
\end{enumerate}

In Fig \ref{fig:garmentexplained} and Table \ref{tab:garmentfeatures} the components and descriptions of garment is shown.



\begin{table}[h]
\centering

\caption{Table showing different components of the garment and their description }
\label{tab:garmentfeatures}
\begin{tabular}{|l|p{0.2\linewidth} | p{0.6\linewidth} |}
\hline
\textbf{Sl} & \textbf{Component}        & \textbf{Description}                                                                                                                                      \\ \hline
1           & Pocket                    & This inside pocket is for placing the heatpad                                                                                                             \\ \hline
2           & Wing Pocket               & This is meant to accommodate the extra portion of the wings in order to minimize the exposure of wings to the inner thigh area reducing chances of rashes \\ \hline
3           & Gusset                    & This is the 3 layer leak gusset which prevents any leakage or stain marks to the external garment                                                         \\ \hline
4           & Placement of sanitary pad & This is how the placement of sanitary pad could be placed by the user                                                                                     \\ \hline
\end{tabular}
\end{table}

\subsection{IoT Integration}
The IoT system of MIMA was designed to provide safe, controllable and distributed heating around the waistband. We developed a Bluetooth-controlled heating system and an Android app to achieve this. For the former, a Heat-Pad (dimensions=180mmx10mm) was fabricated out of Nichrome wire and heat-resistant Kepton\textsuperscript{\textregistered} Polyamide tape. This is tethered to an external power and control module herein referred to as MIMA Control Module.

\begin{figure}
     \centering
      \begin{subfigure}[H]{0.3\linewidth}
         \centering
         \includegraphics[width=\linewidth]{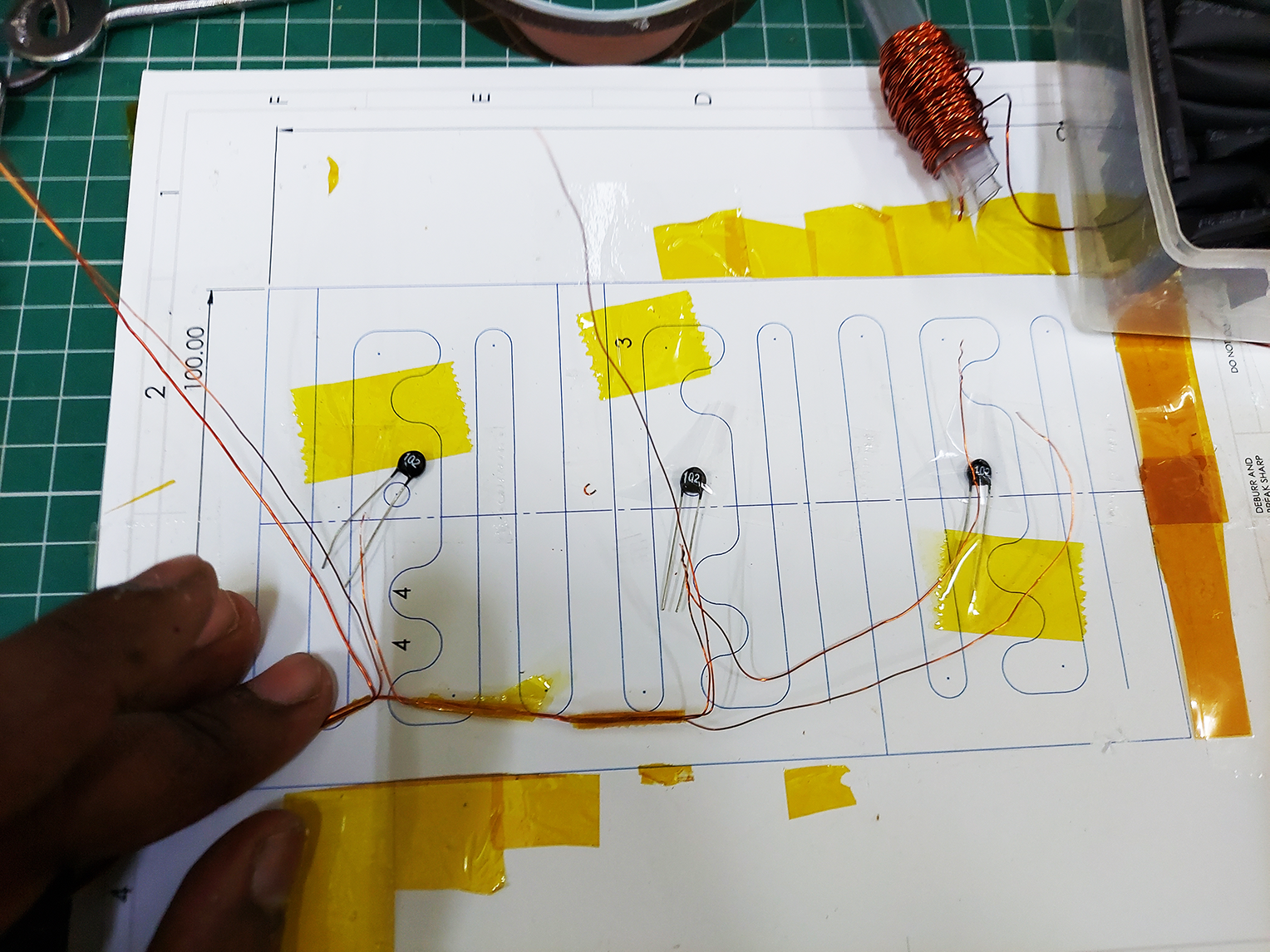}
         \caption{Heat-Pad construction}
         \label{fig:HeatpadConstruction}
     \end{subfigure}
     \hfill
     \begin{subfigure}{0.3\linewidth}
         \centering
         \includegraphics[width=\linewidth]{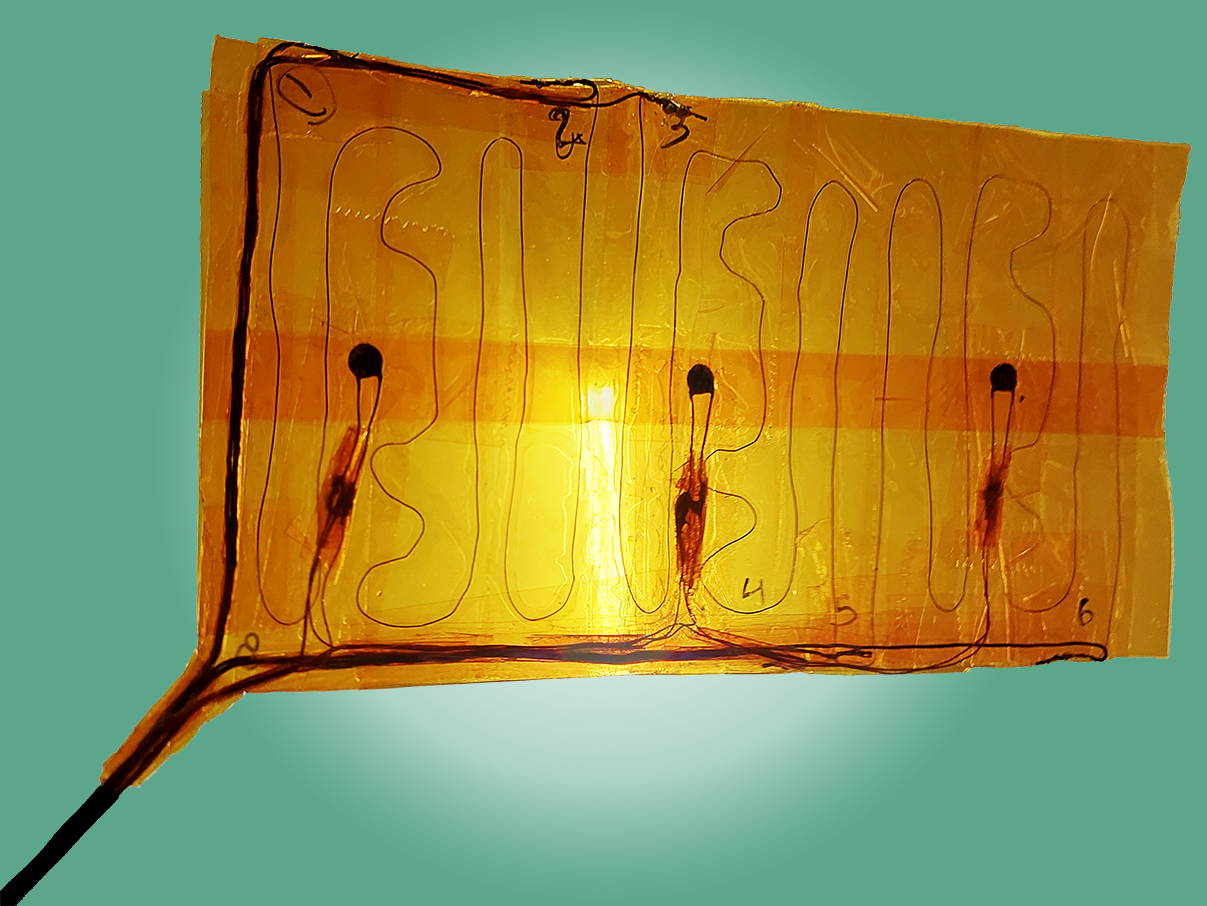}
         \caption{Heat-Pad close-up}
         \label{fig:HeatpadCloseup}
     \end{subfigure}
     \hfill
     \begin{subfigure}[H]{0.3\linewidth}
         \centering
         \includegraphics[width=\linewidth]{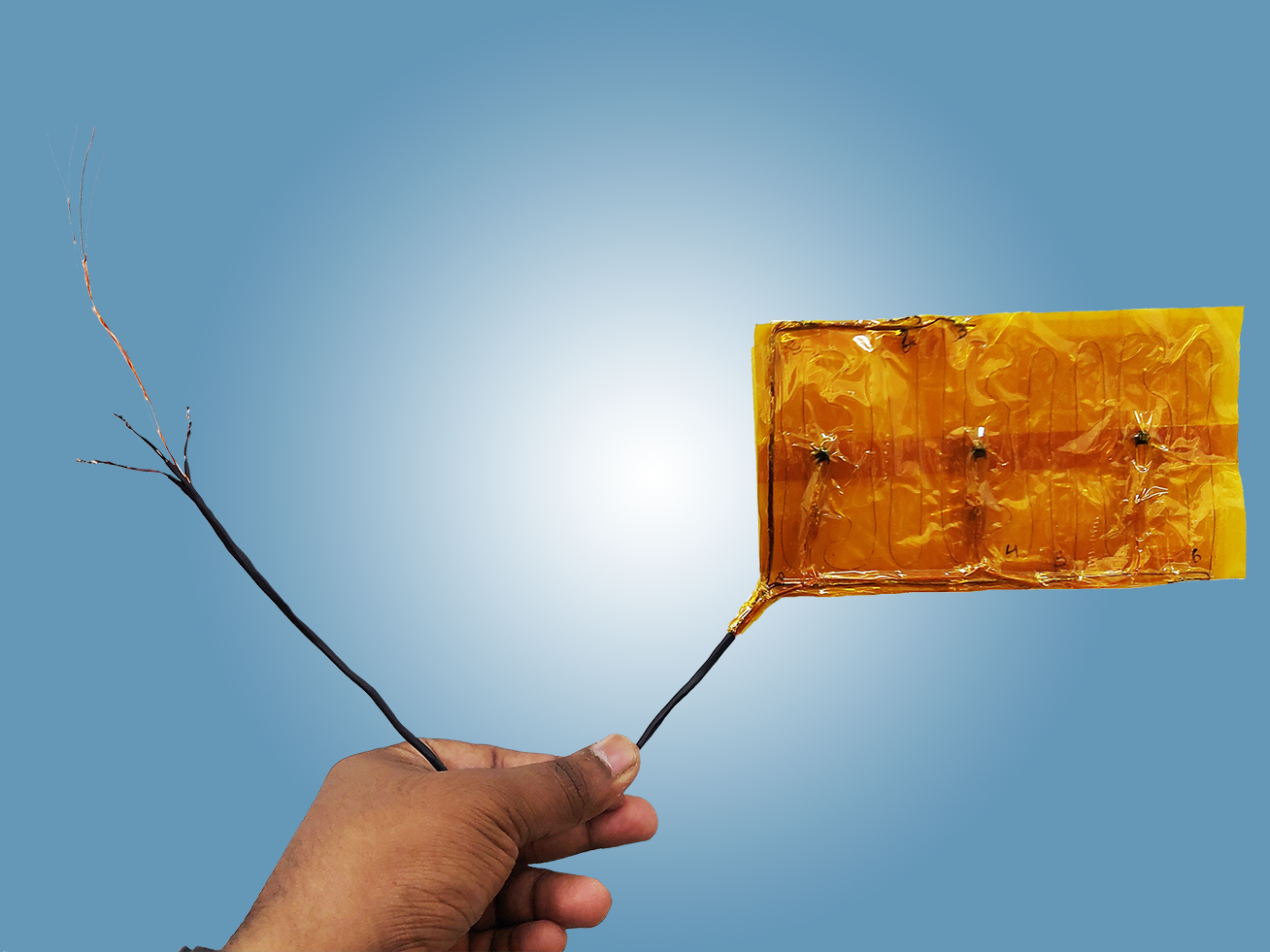}
         \caption{Heat-Pad wiring harness}
         \label{fig: HeatpadWiring}
     \end{subfigure}
     \hfill
    
     \begin{subfigure}[H]{0.45\linewidth}
         \centering
         \includegraphics[width=\linewidth]{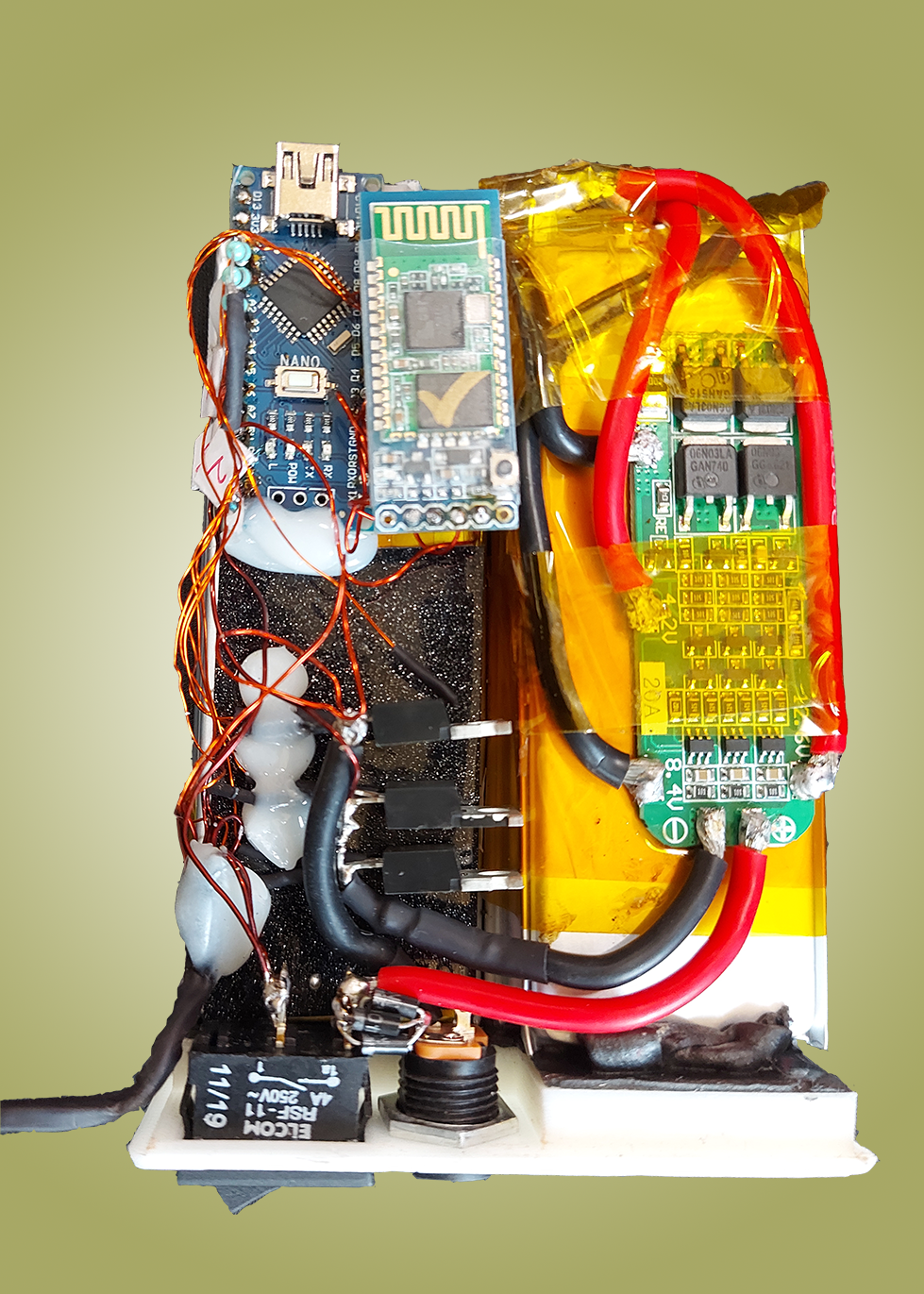}
         \caption{MIMA Controller}
         \label{fig:MIMAController}
     \end{subfigure}
     \hfill
     \begin{subfigure}[H]{0.45\linewidth}
         \centering
         \includegraphics[width=\linewidth]{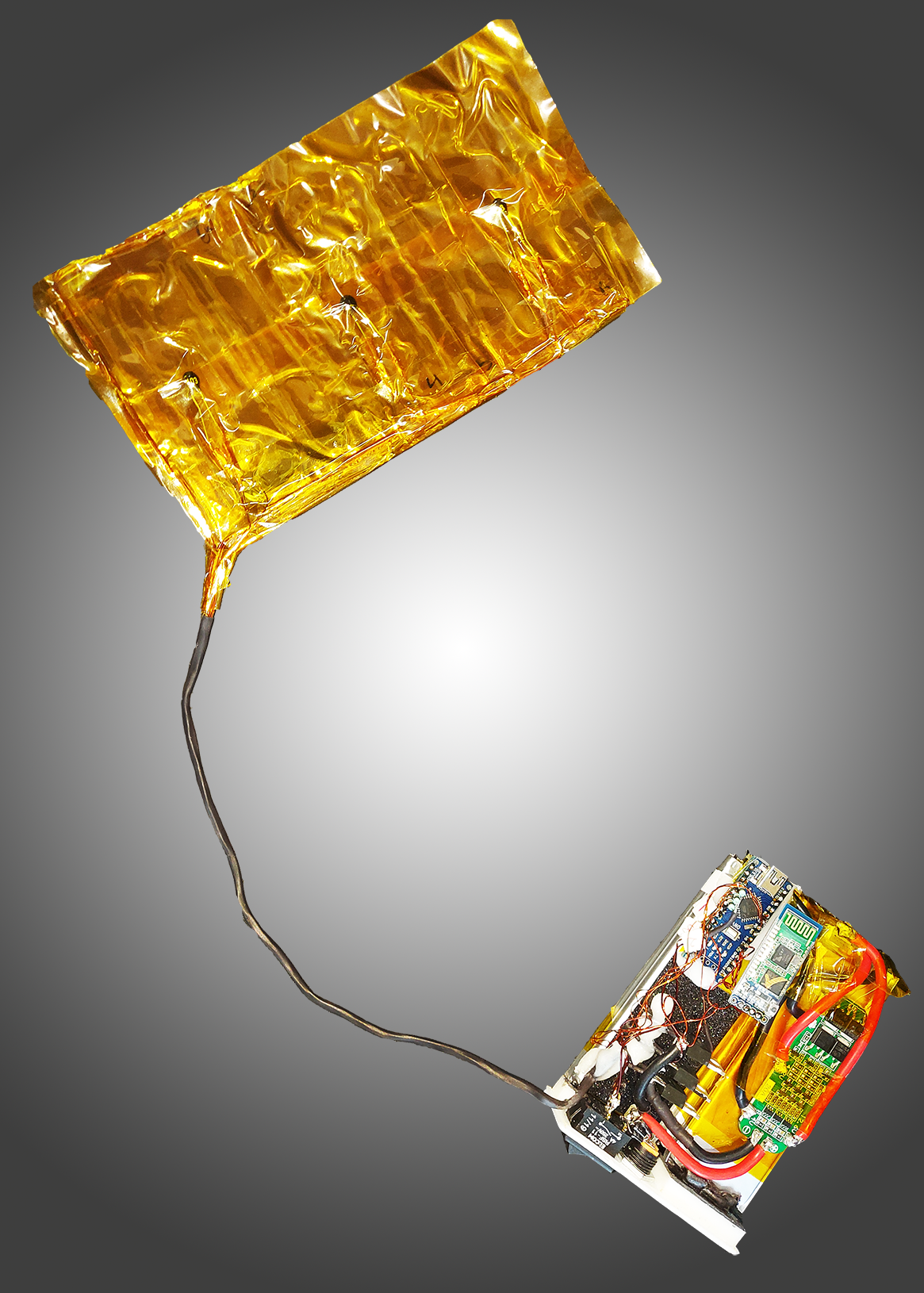}
         \caption{Device Assembly}
         \label{fig:Final Prototype}
     \end{subfigure}
     \hfill
    
     \caption{MIMA IoT Module Prototype Construction}
        \label{fig:MIMA Module}
      

\end{figure}

\begin{figure}[htp]
     \centering
     \includegraphics[width=0.45\linewidth]{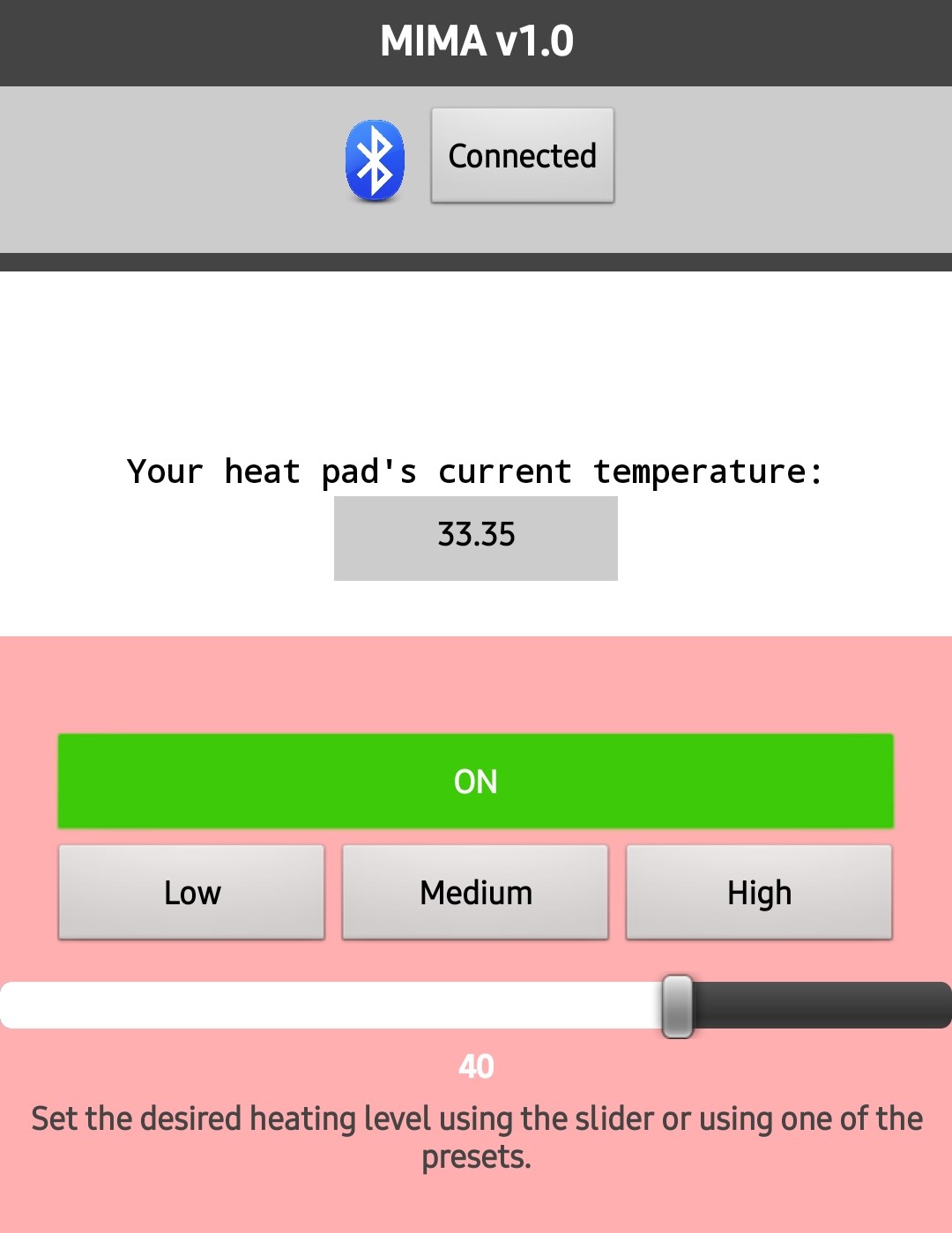}
    \caption{Screenshot of the Android app interface}
    \label{fig:screenshot}
    
\end{figure}

Materials used for the construction of MIMA Control Module and Heat-Pad: 
\begin{itemize}
    \item Arduino Nano
    \item HC-05 Bluetooth Module
    \item IRF540N MOS-FETs
    \item 32 AWG Nichrome wire (Used as heating element)
    \item NTC-103 Thermal Resistors (Used as temperature sensor)
    \item Kepton\textsuperscript{\textregistered} Polyamide tape
    \item 3.7 volt 25c Li-Polymer Cells (Later fabricated into a 3S configuration)
    \item 3S 20Amp BMS Board 
    \item  Enamel coated copper wire 23, 25, 32 AWG
    \item Other small supplies such as resistors, Power toggle button, Charging port etc. 
\end{itemize}
The MIMA Control Module connects with the Android app \ref{fig:screenshot} for sending temperature sensor values to the app as well as to receive temperature control from the app. In our prototype, The Power and Control module is connected with the heat-pad through a thin, flexible cable made out of 10 strands of enamel coated copper wires of suitable gauges for transmitting power (to the Heat-Pad) and for the connection of the Heat-Pad's temperature sensors with the MIMA Control Module. The fabrication of the Heat-Pad and control module is depicted in the figure \ref{fig:MIMA Module} 

\begin{figure}[h]
     \centering
     \begin{subfigure}[H]{0.24\textwidth}
        \centering
        \includegraphics[width=\linewidth]{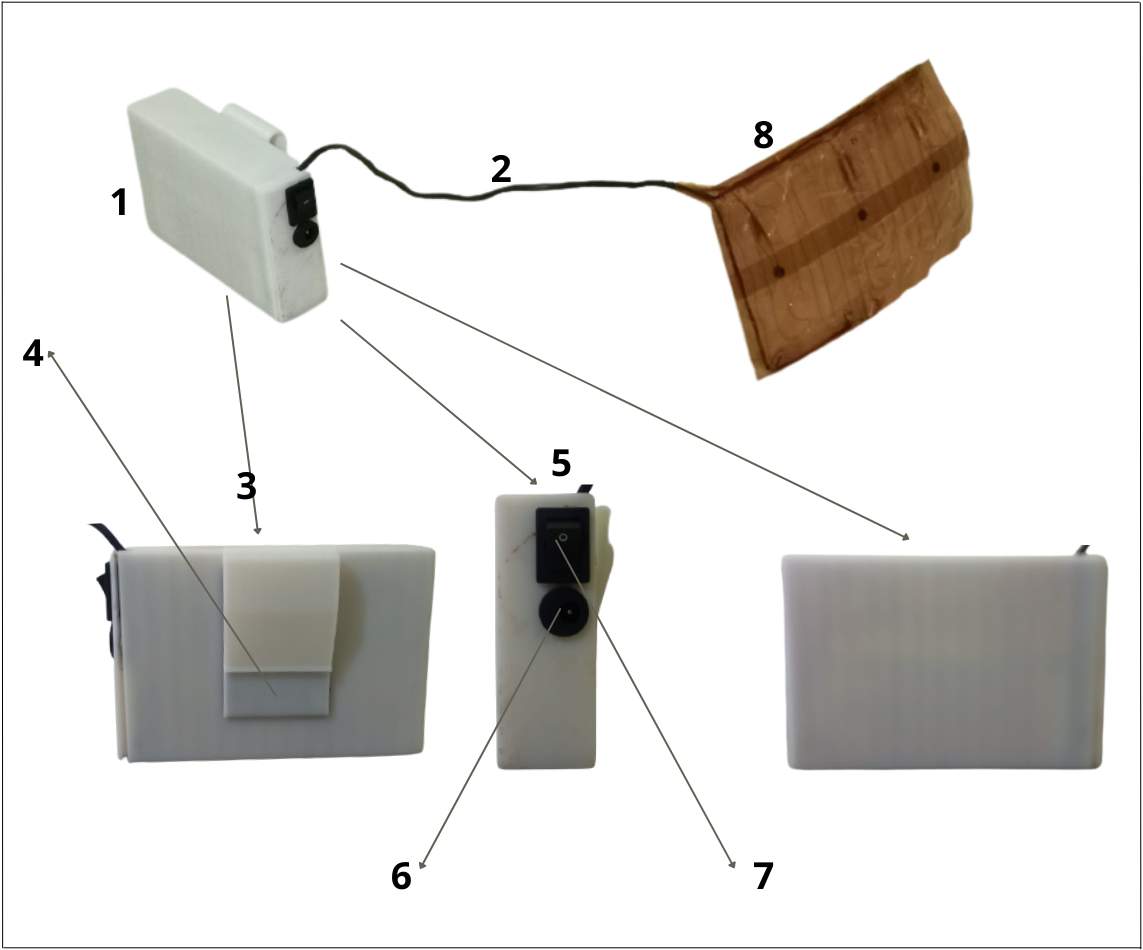}
        \caption{Control module and Heatpad}
        \label{fig:circuit}
     \end{subfigure}
     \hfill
     \begin{subfigure}[H]{0.24\textwidth}
            \centering
            \includegraphics[width=\linewidth]{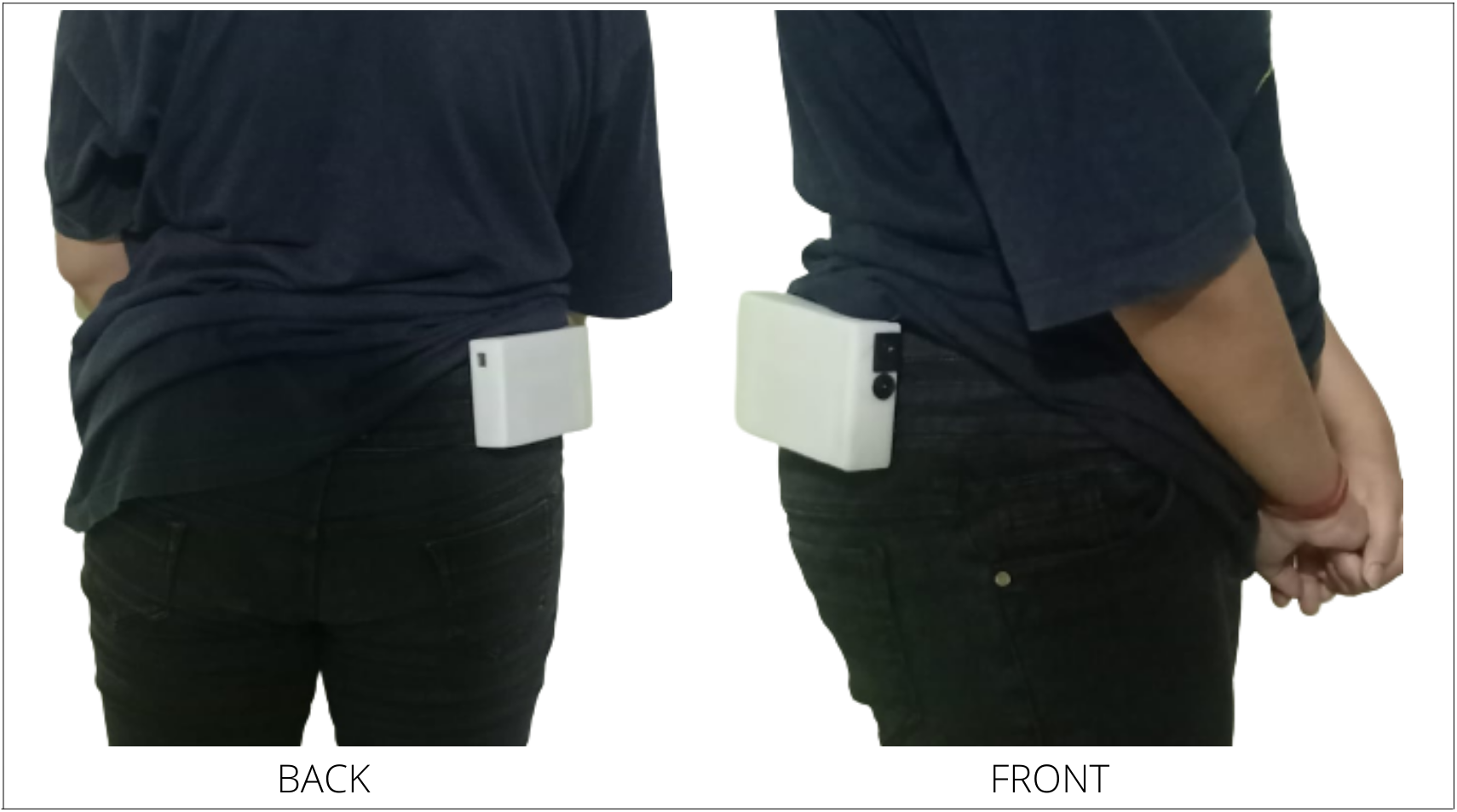}
            \caption{Control module with User}
            \label{fig:user}
     \end{subfigure}
     \hfill
        \caption{Control Module}
        \label{fig:controlmodule}
\end{figure}

The Heat-Pad is divided into three zones (left, middle, and right), and each zone is regulated individually through dedicated thermistor sensors. Based on the input temperature set from the app, the three zones are controlled individually by Arduino through three MOS-FETs. The controller module in our prototype is enclosed within a 3D printed case \ref{fig:circuit}.

We took proper care while writing the firmware to make the system safe by employing multiple checks - some of which are listed below. We have made the code base for MIMA open source, and it is available on our GitHub Repository: \href{https://github.com/smlab-niser/mima2022}{https://github.com/smlab-niser/mima2022}
\begin{itemize}
    \item The standard deviation of temperature sensor readings must be under 2.5 - This is to make sure all three zones are at similar temperatures (which should normally be the case)
    \item The heating of each coil is capped at 55 degrees Celsius for safe operation.
    \item All anomalies are reported back to the mobile app, and for the safe operation of the system, We have configured it so that the module must be connected with the mobile app at all times to be functional.
    \item The Bluetooth module of each of these IoT units is locked using a unique password, ensuring that only the owner will be able to pair to the module using their Android phone.
\end{itemize} 

\begin{table}[h]
\caption{Table showing different components of control module and heatpad}\label{tab:controlmodule}
\centering
\begin{tabular}{|l|p{0.2\linewidth}| p{0.6\linewidth}|}
\hline
\# & \textbf{Component} & \textbf{Description}                                         \\ \hline
1     & Control module        & Contains the circuit components compactly                    \\ \hline
2     & Cable              & Connects the control module to the heat pad                     \\ \hline
3     & Back View          & Consists of the clip                                         \\ \hline
4     & Clip               & It helps to clip the control module to the waist of bottom wear \\ \hline
5     & Side view          & Consists of switch and charging port                         \\ \hline
6     & Charging port      & Used for charging the battery of the control module             \\ \hline
7 & Switch & Used to switch on the system as well as the bluetooth and can also be used to shut down the system in case of application failure \\ \hline
8     & Heatpad            & Placed inside the heatpad pocket used for heat therapy       \\ \hline
\end{tabular}
\end{table}

\section{Result and Analysis}
The Heat-Pad of MIMA gets Incorporated inside an internal pocket in the MIMA period pants. The control module is clipped at the waist of the bottom wear the woman is wearing. Currently, the weight of the control module is around 200gms with the dimension 10.5cm x 7cm x 2.5cm. MIMA was tried by the participants, and their feedback was recorded, and the discussions were done regarding its comfort and overall experience of it. The feedback has been summarised in Table \ref{tab:feedback}.  

\begin{figure}[htp]
     \centering
     \includegraphics[width=\linewidth]{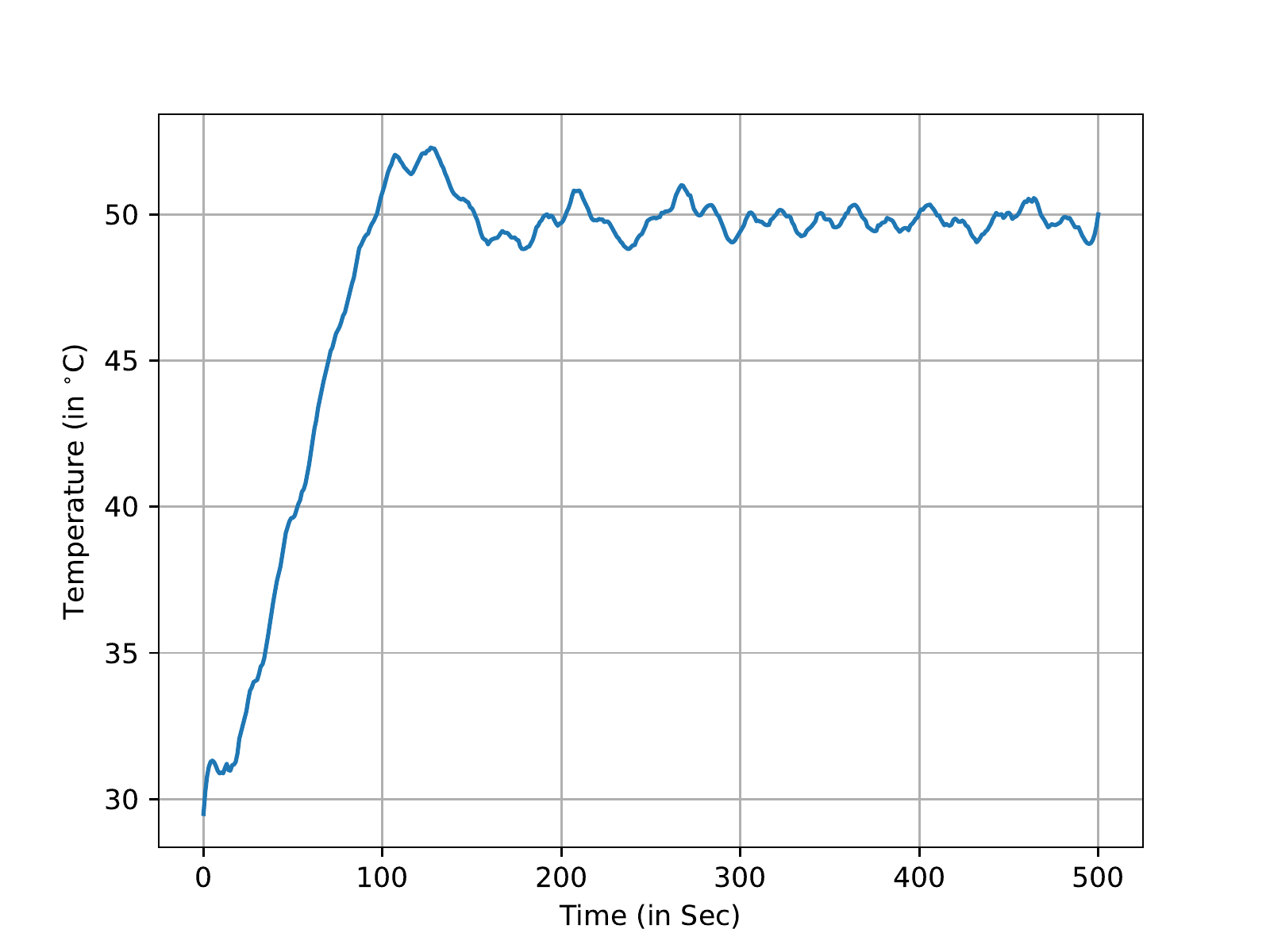}
    \caption{Heating curve of MIMAs IoT Module}
    \label{fig:heatingcurve}
    
\end{figure}

\textbf{Experimental Statistics from MIMAs IoT Module:} The following graph represents the heating curve of MIMAs Heat-Pad\ref{fig:heatingcurve}.
We collected readings at a frequency of 1Hz in a room with an ambient temperature of 30 degrees Celsius. System was powered on and set at the "High" setting of the Android App, which sets the target temperature of the Heat-Pad at 50 degrees Celsius. As depicted, the target temperature was achieved within 95 Seconds. The device was able to maintain a consistent temperature throughout the duration of our experimental run with an average deviation of +/-0.605 degrees Celsius. As it stands, The MIMA Control Module and Heat-Pad module are powered by a 2200 mAh, 12 Volt battery which powers the unit for a constant run at "High" set for approximately an hour. The exact battery backup varies on multiple factors, such as thermal insulation (what sort of external clothing is worn on top of MIMA Period Pants), ambient temperature. Intermittent usage for stretches of 5-10  minutes is advised, and the current battery system is sufficient to power the module through a working day of 6-8 Hours.

\begin{table}[h]
\caption{Feedback on MIMA} 
\label{tab:feedback}

\centering
\begin{tabular}{|p{0.18\linewidth}|p{0.11\linewidth}|p{0.3\linewidth}|}
\hline
\textbf{\begin{tabular}[c]{@{}l@{}}Feedback \\ Criteria\end{tabular}} & \multicolumn{1}{l|}{\textbf{Responses}} & \textbf{\begin{tabular}[c]{@{}l@{}}Representative \\ Statement\end{tabular}} \\ \hline
Demography & \multicolumn{2}{l|}{\begin{tabular}[c]{@{}l@{}}MIMA was tried by participants from age group of \\ 17- 42 years and have helped us with their responses\end{tabular}} \\ \hline
\begin{tabular}[c]{@{}l@{}}Comfort of \\ MIMA\end{tabular} & \multicolumn{1}{l|}{\begin{tabular}[c]{@{}l@{}}60\% of the participants \\ were overwhelmed.\end{tabular}} & \begin{tabular}[c]{@{}l@{}}"It feels very pleasant \\ and cozy!"\end{tabular} \\ \hline
\begin{tabular}[c]{@{}l@{}}Design \& \\ Shade\end{tabular} & \multicolumn{1}{l|}{\begin{tabular}[c]{@{}l@{}}MIMA 's design(80)\%) \\ \& shade(100\%) was very\\ much appreciated \\ and preferred.\end{tabular}} & \begin{tabular}[c]{@{}l@{}}"The shade makes it \\ very easy \& confident \\ for me to carry."\end{tabular} \\ \hline
\begin{tabular}[c]{@{}l@{}}Comfortability \\ of wings.\end{tabular} & \multicolumn{1}{l|}{\begin{tabular}[c]{@{}l@{}}Around 80\% of \\ participants found the \\ wings very helpful \\ and comfortable.\end{tabular}} & \begin{tabular}[c]{@{}l@{}}"I am glad that \\ someone finally \\ thought about this."\end{tabular} \\ \hline
\begin{tabular}[c]{@{}l@{}}Heat-pad \\ comfortability\end{tabular} & \multicolumn{1}{l|}{\begin{tabular}[c]{@{}l@{}}Around 60\% of the \\ participants appreciate \\ the comfort of the heat\\ -pad and most of them \\ prefer medium range of heat\end{tabular}} & \begin{tabular}[c]{@{}l@{}}"Now that I can \\ carry the heat-pad \\ so easily, I don't \\ have to skip a day \\ at workplace"\end{tabular} \\ \hline
\begin{tabular}[c]{@{}l@{}}Ease of \\ clipping\end{tabular} & \multicolumn{1}{l|}{\begin{tabular}[c]{@{}l@{}}Though 70\% of the \\ participants find the \\ clipping very easy, 20\% \\ still took a little bit of \\ time to get it done.\end{tabular}} & "It's pretty easy" \\ \hline
\begin{tabular}[c]{@{}l@{}}Over all feel \\ of the Intimate \\ wear\end{tabular} & \multicolumn{1}{l|}{\begin{tabular}[c]{@{}l@{}}Around 60\% of the \\ women find the over all \\ feel, very comfortable\end{tabular}} & \begin{tabular}[c]{@{}l@{}}"MIMA is different \\ yet feels beautiful"\end{tabular} \\ \hline
App UI/UX & \multicolumn{1}{l|}{\begin{tabular}[c]{@{}l@{}}All the  users find the \\ app very user-friendly\end{tabular}} & \begin{tabular}[c]{@{}l@{}}"I can find whatever I \\ need, it's a cakewalk"\end{tabular} \\ \hline
\begin{tabular}[c]{@{}l@{}}Placing of \\ heat-pad\\ in the pocket\end{tabular} & \multicolumn{1}{l|}{\begin{tabular}[c]{@{}l@{}}60\% of the participants \\ feel that the placing is \\ very easy.\end{tabular}} & \begin{tabular}[c]{@{}l@{}}"Its for the first time, \\ I doing it and I get it, \\ I'm quick at it."\end{tabular} \\ \hline
\end{tabular}
\end{table}

\section{Conclusion and Future Work}

MIMA presents intimate wear that caters to menstrual cramps and rash problems and provides a stain-free menstrual cycle. It is a garment with added functionalities like a heat pad pocket and sanitary pad wing-shaped pocket. We have provided range control of the heating pad to our users through an app that links to the control module through Bluetooth. A few participants tried MIMA, and it was really appreciated for the multiple comfort features and the easy heat therapy it provides during menstruation.

Our future works mainly concentrate on improvising the all-around user experience from garment comfort to application, so we are working on building a lighter, more compact, and detachable control module with easy connectivity and cost-effective. We want to contribute to the expanded scope by developing a product with similar features for the age group of 9+ years as we infer that there is an early onset of menarche among girls, and their needs are not specifically catered to.

\section*{Author Contributions}


Shreya and Amish contributed to the initial ideation, primary and secondary research, along with MIMA's design and development.
Jyothish developed MIMA's IoT system; this the includes design and fabrication of the controller unit, heat-pad as well as the development of the device firmware and Android application. Sulagna and Subhankar contributed to oversight and leadership responsibility for the research activity planning and execution as well as critical review and editing. 


\bibliographystyle{IEEEtran}
\bibliography{manuscript}

\end{document}